\documentstyle[12pt,epsfig]{article}

\def \00 {$0^+\to 0^+$}

\def \EWSB {Electroweak  scale breaking}

\def \SCM {Standard Cosmology Model}

\def \QED {quantum electrodynamics}
\def \SM {Standard Model}

\def  \GUT {Grand Unified Theory}

\def \KK {Kaluza-Klein}

\def  \SS {superstring}

\def \be {\begin{equation}}
\def \ee {\end{equation}}
\def \beq {\begin{eqnarray}}
\def \eeq {\end{eqnarray}}

\def \STR  {Special Theory of Relativity}

\begin{document}

\small

\vspace{0.1in}
\begin{flushright}
{ To the memory of Vladimir Ammosov-}\\
{Friend and Colleague in the Life }\\
{and   in Neutrino Physics}\\
\end{flushright}

\vspace{0.1in}
\begin{centering}

{ THE POSSIBLE SIGNALS }\\
\vspace{0.1in}
{ FROM THE D=6 SPACE-TIME}\\
\vspace{0.1in}

{ Guennady   Volkov}\\
\vspace{0.1in}
{\small
{ \it Petersburg Nuclear Physics Institute}\\
 {\it Gatchina, 188 300 St. Petersburg, Russia}}\\


\begin{abstract}

Based on the exceptional properties of $U(1_{em})$- sterile neutrinos and their possible Majorana-Weyl
space-time nature we continue to discuss our ideas about the existence in \SM \, some  phenomena  or hidden  symmetries
which are related to the extra global dimensions.
We carry on discussing  some open questions of the \SM from the $D=p_s+q_t=6$ space-time geometry.  
  The origin of the three families  could be considered  in 
analogue to the existing of two species- matter-antimatter- in $D=3+1$ Minkowski space-time what  can give the ternary relation for neutrinos
masses. 
Apart from the possible existence of a " hidden neutrino  boost"
 $c_{new}> v_{neut} > c_{light}$  discussed in
 \cite{AV}, \cite{Zichichi}, \cite{LNGS},\cite{CNGS},\cite{Paris},\cite{NDMNP} and a possible indication on $v_{neut}>c_{light}$
 first observed in \cite{OPERA2011},   there could be   some other signatures  of a  $D=6$-vacuum, 
for example,   the possible signals of the
 breaking   $CPT$- invariance  and $Q_{em}$-conservation 
\cite{Paris}, \cite{NDMNP}.
\end{abstract}

\end{centering}

\newpage
\tableofcontents

\section{ Duality between  internal and external space-time symmetries.}
After long searches of decisions so a great number of the  \SM\,
$(Glashow-Salam-Weinberg \,Model$) extensions  realized within the limits of the $D=3+1$-\STR\,  and in numerous  ways of extensions of the internal symmetries of the \SM , physicists  began to introduce also the new concepts in geometry of the world surrounding us which assumes inclusion of new global additional dimensions.

This means that we stand at the beginning of expansion of some concepts of the Special Theory of the Relativity. First of all it concerns to multidimensional $(D> 4)$ to generalization of Lorentz group  and more careful analysis, that role which the velocity of light plays in  this theory and space-time nature of neutrinos. What parameters will define applicability of the subsequent theory of a relativity  there is the most important question in opening of new \STR.

All known physical theories   are described by internal and external space-time symmetries.
To understand these theories  much deeper one should find the correlations between them. Therefore we follow for   the evolution of the main physical theories
looking  for     the correlations  between  internal and external  space-time symmetries : 
\begin{eqnarray}
\begin{array}{||c|c||c|c|c||}\hline
 N &  Phys.Theor.&Ext.Sym&IntSym& SpaTime\\\hline\hline
 1& ClasMech& GalRel    & NewtDyn& R^3\\\hline
2&QED&   STR-SO(3,1)     & U(1_{em})&R^4_{3,1}\\\hline
3& SM&   STR^{*}-SO(3,1) &SU(3)\times SU(2)\times U(1)&R^4_{3,1}1? \\\hline
4& GUT&  STR-SO(3,1)? & SU(5),SO(10),E(6)    &R^4_{3,1}?? \\\hline
5^{1}& HetSS& SO(9,1),N=1SUSY ?& E(8)\times E(8)&R^{10}_{9,1}\\\hline 
5^{2}& ComHetSS &SO(3,1),N=1SUSY?& E(8)\times E(6)& R^4_{3,1}\otimes K_6\\\hline
 \end{array}
 \end{eqnarray}
  We started from the classical mechanics and ended by the $E(8)\times E(8)$ heterotic  \SS \, in $R^{10}_{9,1}$
  and with $E(8)\times E(6)$ in $R^4_{3,1} \times K_6$ space\cite{CHSW}.

 It becomes more and more reasonable to propose
that the Standard Model and its extensions could have  an intrinsic link 
with a more fundamental symmetry, 
than the Cartan-Lie symmetries. Of course, this fundamental symmetry
should generalize the symmetries of the Standard Model, 
since a lot of experimental data confirm its. 

Therefore  a hypothetical symmetry  should be a natural
generalization  of the  Lie symmetries with  stronger 
constraints leading to diminishing the number of free parameters  of the \SM. 
 In principle, in (super)string approach we already have the interesting 
example of the  generalization of finite Lie algebras by an 
infinite-dimensional affine Kac-Moody  algebra with  a central charge.
Superstring theory intrinsically contains  a
number of infinite-dimensional algebraic symmetries, such as the Virasoro
algebra associated with conformal invariance and affine Kac-Moody algebras. 
 On the first  traditional  way to get in \SS \,  the \SM\, at low energy limit
 there should be solved the problem of compactification extra dimensions,
 to get the Minkowski $R^4_{1,3}$ space-time.
On this way it was  discovered  for physics  the very interesting  geometrical objects-Calabi-Yau manifolds, more exactly,  the heterotic $E(8)\times E(8)$ superstring have been compactified on the 6-dimensional Calabi-Yau space $CY_3$, having the $SU(2)$,  $SU(3)$ group of holonomy  \cite{CHSW}, \cite{CF}, \cite{AENV}. 
Lucky happened  that the structure 
of $ CY_2 \subset CY_3 $ spaces are directly connected to
the affine Cartan-Lie symmetries\cite{CF}. 
This link of $CY_2$ spaces with  some algebras from $A_r^{(1)}$, $D_{r}^{(1)}$,  
$E_6^{(1)}$, $E_7^{(1)}$, $E_8^{(1)}$ algebras 
can be explained by   the  crepent 
resolution of ADE-type  singular structures  ${ {\bf C}^2/G}$,
where $G$ is one the five  finite  subgroups  of $SU(2)$.
Thus the\SS\, approach generated  by  ideas of Kaluza-Klein of the link the internal symmetries and the singularities to the cycle structures of the compactified manifolds like $CY_2,CY_3$ \cite{CF},\cite{AENV},\cite{HUN}, \cite{Volk}.

The main argument to look for  new 
symmetry with some extra ordinary properties is that the symmetries linked
to the simple Cartan-Lie groups are not sufficient for a description of 
many parameters and  features of the Standard Model.  
Therefore  a hypothetical symmetry  could be a natural
generalization  of the  Lie symmetries with  stronger 
constraints which could lead to diminishing the number of the free parameters of the \SM.  There could be exist some ways to find these algebras and symmetries. One very traditional way is going from the 
theory of numbers. For example to consider the new complex numbers 
$q^D=1,-1$
$(D=3,4,5,...)$  \cite{Appel},\cite{Dev},{\cite{Fleury}, \cite{Kern},
\cite{Rausch},\cite{LRV},\cite{DV}, \cite{Volk}   to find  the geometrical figures of
defined by equations  
\beq   
z \cdot {\tilde z} \cdot  {\tilde {\tilde z}} \cdot \cdots=1,
\qquad z\in R^D,  \nonumber\\
\eeq
where $R^D$ is the $D$-dimensional Euclidean space.
First one can find the   $D=3,4,...$-dimensional Abelian extensions of $U(1)$-group.
and then to look for the non-Abelian extensions.
Another way is the geometrical way to study the singular structure of 
$CY_3$-, $CY_4$-spaces.

 But also there exists another important way connected to the study  of the external symmetries  connected to the very big extra dimensions.

In \QED \, and in \SM\,  the external symmetries are connected with  the ambient geometry of our  (3+1) space-time. 
These symmetries can be be described by Lorentz group,$SO(3,1)$, having 6-parameters,
6=3-rotations + 3-boosts,  and plus four translations $ x_{\mu}^{\prime}= x_{\mu}+ \xi_{\mu}$.
Also the external symmetries include some discrete symmetries, like, 
$P$, $T$, and $C$ symmetries,
which are unified into  the principal discrete symmetry of the 
SM - $CPT$ \cite{BS},\cite{BLOT},\cite{Ram}.

 It has been already considered a lot of the models extended the \SM \, at very big energy range and, correspondingly, at very small distances,
where it was proposed  that the  \STR  is still correct?!
As one can see from the table the Grand Unified Models \cite{Pati},\cite{Georgi},\cite{Mink},\cite{CHSW}, \cite{AEHN},\cite{VOL}
with 
$SU(5)$,  $SO(10)$, $E(6)$ or $E(8)\times E(6)$- groups 
with the scale of unification about $10^{16}$ GeV, used on this scale the 
external  Lorentz space-time symmetry $SO(3,1)$ and, correspondingly, the \STR ?!   
The transformation  of the special theory of a relativity on these scales could cause the non  observability those channels of the proton decays   predicted by   \GUT s.
Thus  in some \GUT s the questions about the external space-time symmetry  and other  vacuum features often remain opened which  can lead to the wrong experimental predictions.

 In the \SM \,  one can also consider 
 a link between  external and internal global conservation laws. As the most important example one should study the correlation  
between the electric charge conservation and $CPT$ invariance\cite{Paris}:
\be 
CPT -\,invariance\,\,\, \leftrightarrow  \,\,\, Q_{em}- \,conservation.
\ee

The conservation of the electromagnetic charge is the consequence of the
$U(1)$- global symmetry.  There is remarkable link between the charges of proton and electron: 
\be  
|Q_p\, \,+\,\, Q_e| \qquad  \leq  \qquad 10^{-21}q_e\\ 
\ee
The $CPT$ theorem can be proved in any Hermitian  Lorentz symmetric quantum field theory (local)\cite{BLOT}. 
The $CPT$-invariance gives the very important  equality for the masses(time life) of particle and antiparticle:
\be 
m(\Psi)\, =\, m({\bar{\Psi}})\,\,\rightarrow\,  binary \,\,
 CPT \,-\,invariance
\ee
 
We have got a lot of  experimental and theoretical indications that $Q_{em}$-conservation law is valid  in the $D=3+1$ Minkowski space-time.
One can  
Thus in this approach the $CPT$-invariance and $Q_{em}$-conservation law are the prerogatives only for Minkowski $R^4_{3,1}$- space-time where are  valid the $SO(3,1)$  Lorentz group (Poincar$\acute {e}$) symmetry and
$U(1_{em})$ gauge symmetry\cite{AV},\cite{CNGS},\cite{LNGS}, \cite{NDMNP}. This validity of duality between global internal and external symmetries- laws can be checked on some  experiments \cite{Paris}. 
 
 There is the well-known the Poincar$\acute{e}$ duality in algebraic topology   between the Homological group of manifold and cohomological group of the differential forms connected with this manifold.  It  means that if you knows something about the topological structure of the closed but non exact  manifolds (Betti numbers) it is possible to connect them to the  Hodge numbers characterize  the Rham cohomological group of  the closed differential
 forms but non exact differential forms determined on this manifold and opposite.

  One can suggest the following relations  between the topological geometry  and physics:
 \begin{eqnarray} 
 \begin{array}{ccc}
Homology \,\, Manifold \,\,&\Rightarrow \,\,& Space-Time; \\
 Cohomolgy \,\, Forms \,\, &\Rightarrow \,\,  &Fundamental \,\, Particles\\ 
 \end{array}
 \end{eqnarray}

 In physics the analogue of this duality means that the properties of the fundamental particles ( like 
 electron and positron in quantum electrodynamics) give very important information about 
 the structure of the  space-time, for example,  about its dimension.

 Thus, if such a duality 
exists,  processes violating the CPT-invariance should accompanied by 
electromagnetic charge violation too. May be,  this was one of the reasons 
why we did not see some rare decays, such as proton decay. In this case, the 
idea of grand unification is not enough to predict  baryon or lepton- 
violating processes. Also a similar problem could be happened with  searches for 
the rare flavour-changing processes in the channels, such as $\mu \rightarrow e + \gamma$,
$\mu \rightarrow 3e$ etc (if the observed broken family symmetry has an external space-time origin)
Our idea of looking such processes is related to the large extra dimensions.

Thus one can say that to achieve in the \SS \,  the progress it was shown   one the ways how to solve  the very important question i.e.,
the  internal symmetries can be got by the geometrical way,  from the study the singularities 
of the  compactified $CY_n$ spaces. This way begun from the old idea of \KK when the
$C/Z_2$ singularity produces the $U(1_{em})=S^1$ symmetry. The compactification on $CY_3$ spaces could expand the \KK \, case at very big class of  simple   Cartan-Killing-Lee algebras \cite{CF}.

There is another  very perspective way to consider  non shrink
global extra dimensions \cite{ADD},\cite{Ran}, \cite{Rub}. There  is 
immediately  appears    very difficult problem - what is the \STR in $D>4$? What could be the Lorentz group extension?  In \SS\, approach there was used the $D-$ dimensional non compact group Lorentz with
quadratic $(p,q)$-metrics $( +...+-,...,-)$?!  But we don,t know how to get the point limit and what there is no examples of the renormalizable
quantum field theory in $D>4$ ( except $\phi^3$ in $D=6$?)\cite{BS}. 

The main experience  what we have got from \KK, \SS/D- brane approaches and from the study of the structure of $CY_n$-manifolds
is that we begun to understand  the role  of the compact  small  and non compact  large dimensions in constructing 
the theories including the internal or external space-time symmetries. We can suppose that  the non compact  large dimensions are related to the extra dimensional space-time symmetries of our ambient world. The compact small dimensions are connected to the origin of the internal symmetries.   For the \SM \, this  circumstance was very important since we thought that the problem of three neutrino specie could be solved by including in our $D=3+1$ space-time  the some global  non compact  dimensions\cite{AV},\cite{Paris}. So we begun  to think that the appearance three families  is related to the big extra dimensions like it was happened with two "families", particle-antiparticle, what was explained by Dirac in his relativistic equations  in the $D=3+1$- Minkowski space-time.  Earlier a lot of publications has been devoted to the possibility to solve the three family problems  through the introduction  the internal gauge symmetries
\cite{AEHN},\cite{VOL}.  But inside the three family problem there is  the neutrino problems: 
the the Dirac-Weyl-Majorana nature \cite{Dirac},\cite{Weyl},\cite{Majorana}, their masses  and etc. We plan to consider the 3-neutrino problems by introducing the
the two extra "non-compact" large dimensions and considering the corresponding geometry using the procedure of ternary complexification of Euclidean  $R^3$ space and its extensions into the higher dimensions $D \geq 3$.
Note that all the previous geometrical ideas have been linked to the two dimensional Euclidean, Lobachevski, Riemann spaces and its the  $D =2,3,4,...$ hyper surfaces extensions with the   symmetry-invariant quadratic metrics \cite{Dub}.

 In $D=6$ we can consider some cubic and quartic geometrical surfaces, the tetrahedral Pythagoras theorem, the $3\times 3$
matrices an analogue of the Pauli matrices, n-ary complex analysis in $R^D$ \cite{Appel},\cite{Dev},\cite{Fleury},\cite{Kern}, \cite{Rausch},\cite{DV}, \cite{ LRV},\cite{Volk} .
which give us some new interesting geometrical spaces having some
applications \SM and in \SCM.

\section{The $D=(3+1)$ $STR$  and the  space-time structure of neutrinos}

 It can be distinguished the  main  principles  of \STR as
the principle of relativity, principle of maximal velocity and  the symmetry of the electromagnetic vacuum structure which has been  determined by the 4- dimensional  space-time Lorentz $SO(3,1)$- group. The principle of maximality velocity  can be formulated  like   the maximal velocity of the material objects  cannot be more  than the light velocity in the vacuum.  The vacuum  structure will be very important for the next discussions since
the  notion of the vacuum should be defined correctly  and  it  can depend on  some extra conditions of our ambient space-time.    One can think that,  if our visible world is determined only by the electromagnetic vacuum,  the \STR   should be correct.  But if the observed until now the D=1+3 dimensional world is only a subspace of the higher dimensional space there  emerge the requirements to check the principles of STR on the experiments.  It the case when one transforms  the geometrical properties  of the vacuum structure by embedding the additional dimensions, it is the most plausible that the  principle of the maximal velocity should be crucially  changed i.e. there could exist some specific particles  propagated faster than light ( neutrinos, dark matter "fermions")
\cite{AV},\cite{LNGS},\cite{Zichichi},\cite{Paris}, \cite{OPERA2011}, \cite{MINOS2007}, \cite{Kal}. 
 
 To realize the program to  embed the  "visible" $SO(3,1)$- ambient world into higher dimensional space-time with large non compact extra dimensions  one should  take in mind  all achievements 
 of $D=3+1$ approach in a new multidimensional \STR. 
   Our interests  are connected to the  checking  the origin of some fundamental constants characterized the \STR\, understanding of which 
   could give the progress in solving some problems  like as   
 3-neutrino species, three generations, the fermion and $W,Z$ bosons masses.
Now we  do accent on some peculiarities of \STR\,  Lorentz group  and its spinor representations\cite{BS},\cite{BLOT},\cite{Ram},\cite{Dub}.

So  the Galileo's transformations should be changed by  Lorentz's transformations($ \mu=\{0,i=1,2,3\}$):
\be  
x^{{\mu}\prime}=\Lambda^{\mu}_{\nu}x^{\nu}
\ee
what conserve in Minkowski  space $R^4_1$ the following space-time interval:
\be 
ds^2=g_{\mu\nu}dx^{\mu}dx^{\nu}=c^2 (dt)^2-(dx^1)^2-(dx^2)^2-(dx^3)^2,
\ee
where it was suggested $x_0=ct$ and the Lorentz metrics is :

\be  
g_{\mu\,\nu}\,=\,g^{\mu\,\nu}\,=\,
diag (1,-1,-1,-1) \nonumber\\
\ee

A Lorentz transformation of $ R^4_1 \rightarrow R^4_1$
is a linear map ${\Lambda}$ with
\begin{eqnarray}
(x^{\prime} \cdot x^{\prime} )=( \Lambda x \cdot \,\Lambda x)\,=\,(x\cdot\,x)\,&=&
\,g_{\mu\,\nu}\cdot x^{\mu} x^{\nu} \nonumber\\
\Lambda^{\mu}_{\rho} \cdot g_{\mu\,\nu} \cdot \Lambda^{\nu}_{\tau}\,&=&\, g_{\rho\,\tau}.
\end{eqnarray}

 In STR the speed of light is identical in all inertial systems
i.e. for two coordinate systems $\{t,x^1,x^2,x^3\}$ and $\{t^{\prime},(x^1)^{\prime}, (x^2)^{\prime}, (x^3)^{\prime}\}$ the corresponding intervals must be equal:
\be   
 c^2t^2-(x^1)^2-(x^2)^2-(x^3)^2= c^2 {t^{\prime}}^2 -{(x^1)^{\prime}}^2-{(x^2)^{\prime}}^2-{(x^3)^{\prime}}^2 .
\ee
For  the world line of the material particle in Minkowski space-time $R^4_1$
\be   
x^0=ct,\,\, x^1=x^1(t),\,\,x^2=x^2(t),\,\,x^3=x^3(t),
\ee
one can consider the space vector velocity 
\be   
v=\{\dot x^1,\,\,\dot x^2,\,\,\dot x^3 \}.
\ee
Then the principle of maximality velocity  can be formulated  like   the maximal velocity of the material objects 
 cannot be more  than the light velocity in the vacuum, i.e. 
\be
 |v| \leq c.
\ee

The Lorentz group ${ L=O(3,\,1)}$ consists of 4 connected 
components depending on the sign of the determinant ${ det L=\pm\,1}$
and the magnitudes  of the ${ L^0_0}$-components, ${ L^0_0\geq 1}$
or ${ L^0_0\leq 1}$.  Any ${ \Lambda \in { L_+^{\uparrow}}}$
can be got from boost or rotation.

For each ${ x^{\mu}}$ one can construct a complex Hermitian 
${ 2 \times 2}$ matrices 
\begin{eqnarray}
\sigma(x)\,\,&=&\,x^{\mu} \cdot \sigma_{\mu}\,\,\,\rightarrow
det {\sigma(x)}\,\,=\,(x^{\mu} \cdot x_{\mu}), \nonumber\\
\end{eqnarray}
where ${ \sigma_i}$ and ${ \sigma_0}$ are three Pauli matrices
and ${ 2 \times 2}$ unit matrix, respectively.
The transformation of ${\sigma(x)}$ with a complex
matrix ${ {\hat A}\in SL(2C)}$ , ${ det {\hat A}=1}$:
\beq
\sigma(x)\, \Rightarrow \, \sigma(y)\,=\, \hat {A}\cdot \sigma(x)\cdot {\hat A}^{+},
\eeq
conserves the norm of $\sigma(x)$, {\it i.e.} $det \sigma(x)$.
The mapping : 
${\hat A \rightarrow L_A}$  is a group homomorphism from $ SL(2C)$
onto ${ L_{+}^{\uparrow}}$. 

${ SL(2C)}$ is a simply connected Lie group and is the universal 
covering of ${ L_{+}^{\uparrow}}$.

Any ${ {\hat A}\in SL(2C)}$  can be written as
\begin{eqnarray}
{\hat A}\,=\,{\hat B} \cdot \hat J, 
\end{eqnarray}
where the Hermitian matrix $\hat B$ and the unitary matrix $ \hat J$
determine the  boost and the 3-space rotation, respectively:
\begin{eqnarray}
\hat B\,&\,=\,&\, \exp{[\frac{1}{2}\omega \cdot  (\vec{n} \cdot \vec {\sigma})]},\nonumber\\
\hat J\,&\,=\,&\, \exp{[-i \, \frac{1}{2} \varphi (\vec{n} \cdot \vec {\sigma})]}.
\end{eqnarray}

\
The matrix ${L_{\hat J}}$ is a pure rotation around the axis ${ \vec{n}}$
through an angle ${ \varphi}$.  The operator ${ L_{\hat B}}$ is a pure ${ \vec{n}}$-boost with a velocity
\begin{equation}
\frac{\vec{v}}{c}\,=\, \vec {\beta}=\,\tanh{(\omega)} \cdot \vec{n}.
\end{equation}
If the system $\{x^{\prime}\}$ is moving with respect to the system $\{x\}$ along 
the axis with the speed $v$ the boost transformation takes the next form:
\beq 
(ct)&=&(ct^{\prime})\cosh \omega  + x^{\prime}_1\sinh\omega, \\
x_1&=& (ct^{\prime})\sinh \omega +x_1^{\prime}\cosh\omega,\\
x_2&=&x_2^{\prime},\\
x_3&=&x_3^{\prime}. \\
\eeq
where 
\beq 
\tanh \omega=\frac{v}{c}=\beta,\,\,\gamma = \frac{1}{\sqrt{1-\beta^2}}\\
\cosh \omega=\gamma,\,\,\,\, \sinh\omega=\beta \cdot \gamma
\eeq
 For the energy-momentum  4-Lorentz vector one can get the following Lorentz invariant expression:
\be  
E^2 - \vec p^2 \cdot c^2= m_0^2 \cdot c^4, 
\ee 
where $E$ and $\vec p$ are the energy and the 3-dimensional momentum vector, respectively.  $m_0$ is the mass parameter in the rest.
 
Thus the Lorentz group transformations are determined by the fundamental constants : light speed in vacuum $c$, the rest mass $m_0$
and six  free parameters: three angles and three-vector $\vec \beta$ with the change region:
\be 
  |\vec \beta| \leq 1.
 \ee   
Time by time there were appeared some ideas how to  extend this region
and to find signals  expending faster than light, for example,
searching for   tachyons.  The other ideas have been initiated by cosmology problems to solve some of them by  proposal that the fundamental constants like $C_{light}$ depends on the time?  And the last  comment is concerning the our space-time.   Our representations about the world change with getting more and more information from astrophysics  and 
our progress in particle physics.
So the extension space-time due to including some additional dimensions
immediately causes the questions about the validity of \STR.

The questions about mass origin of the  particles and their properties  are very important. These questions are very closely linked to the Lorentz representations \cite{BLOT}, \cite{Ram},\cite{Pal}.
 
There are well-known six  Lorentz spin generators,$S^{\mu\nu}$, 
and  the following combinations of them:  

\beq J^i=(J^1,J^2,J^3) &=& (-i S^{23},-iS^{13},-iS^{23}) \nonumber \\
B^i=(B^1,B^2,B^3) &=& (-i S^{01},-iS^{02},-iS^{03})  \nonumber\\
\eeq
obey to the  commutation relations

\beq
&&[J^i,J^j]= i  {\epsilon}_{ijk} J^k,   \nonumber \\
&&[B^i,B^j]=-i  {\epsilon}_{ijk} B^k,   \nonumber\\
&&[J^i,B^j]= i  {\epsilon}_{ijk} B^k.   \nonumber  \\
\eeq

To get the isomorphism of $so(3,1)\equiv sl(2,C)$ algebra to the algebra
of the semi-simple group $SU(2)_L\times SU(2)_R$ it is easy to see from the commutation relations of 
the 6-operators $I^{\pm i}=\frac{1}{2} \cdot  ( J^i \, \pm\,  B^i)$:

\beq
&&[I^{-i},I^{-j}]= \epsilon_{ijk}I^{-k}     \nonumber \\
&&[I^{+i},I^{+j}]=\epsilon_{ijk} I^{+k}\nonumber  \\   
&&[I^{+i},I^{-j}]=0\nonumber\\
\eeq

This circumstance allows to classify the irreducible representations of $SL(2,C)$
by two integer or semi-integer numbers $(m,n)$ of the finite-dimensional
representations of the $SU(2)_L\times SU(2)_R$ group.
These representations one can use for spinor representations of $S^{\mu\nu}$.
The minimal representations of $S^{\mu \nu}$
(except scalar $(0,0)$ representation) are the Weyl spinors
left-handed $(\frac{1}{2},0)$- and right-handed 
$(0,\frac{1}{2})$- representations:
\be 
{\vartheta}_a(x)\in (\frac{1}{2},0), \qquad a=1,2 \nonumber\\
\ee
\be 
{\bar \vartheta}^{\dot a}(x)\in (0,\frac{1}{2}),\qquad {\dot a}=1,2 \nonumber\\
\ee

Two $SU(2)_L$- and $SU(2)_R$-  are related by $P$-parity operation:
\be  
P: \qquad x_0 \rightarrow  x_0,\qquad \vec {x}\, \rightarrow\, - \vec{x} 
\ee
In the terms of generators $\vec J$ and $\vec B$ the $SL(2,C)$
transformations  take the forms:
\beq 
&&S=\exp {i[ \vec \varphi \cdot \vec J\,+\, \vec \omega \cdot \vec B]} \nonumber\\
&&S^P=\exp {i[ \vec \varphi \cdot \vec J\,-\, \vec\omega \cdot \vec B]} \nonumber\\
\eeq

The Weyl spinors  cannot have $m \neq 0$.
The Dirac fermions which are in the representation:
$(\frac{1}{2},0)+(0,\frac{1}{2})$,
 
where
\beq 
\Psi=
\left(
\begin{array}{c}
{\vartheta}_a \\
{\bar \eta}^{\dot a}\\
\end{array}
\right )
\eeq

In Majorana  basis  the Dirac matrices are pure imaginary 
$({\gamma}^{\mu})^{\star}=-{\gamma}^{\mu}$:
\beq
\gamma^0=
\left(\begin{array}{cc}
0 & \sigma^2 \\
\sigma^2& 0\\
\end{array}
\right ),\,
\gamma^1=
\left(\begin{array}{cc}
i \sigma^3& 0 \\
0 & i\sigma^3\\
\end{array}
\right ),\,
\gamma^2=
\left(\begin{array}{cc}
0 & -\sigma^2 \\
\sigma^2& 0\\
\end{array}
\right ),\,
\gamma^3=
\left(\begin{array}{cc}
-i \sigma^1&0 \\
0 & -\sigma^1\\
\end{array}
\right ).
\eeq
 In this basis the generators of Lorentz transformations $S^{\mu\nu}$
 are real and the fermion can be also real 
\beq  
\Psi \,=\,{\Psi}^{\star}
\eeq 
This condition is preserved under Lorentz transformation.
The operation of the complex conjugation  can be extended for a general Dirac basis taking only
\beq  
(\gamma^0)^{\dagger}=\gamma^0, \qquad {\gamma^i}^{\dagger}=-\gamma^i. 
\eeq
Then  it was introduced the charge conjugation operator 
$ C=4\times 4$- matrix with the following conditions:
\beq  
C^{\dagger} C=1, \qquad C^{\dagger} {\gamma}^{\mu} C = 
- ({\gamma}^{\mu})^{\star}. \nonumber\\ 
\eeq

 Under action of the charge conjugation Dirac fermion transforms into
\be 
{\Psi}^{C}=\hat C {\Psi}^{\star} \nonumber\\ 
\ee
The 
The charge conjugated fermion ${\Psi}^{C}$ transforms by Lorentz 
covariant way and also satisfies to Dirac equation if the $\Psi$  satisfies. Thus the condition of the Majorana fermion  reality is
\be 
\Psi^{(C)}=\Psi 
\ee
which is Lorentz invariant. In Majorana basis this condition reduces to
$C=1$ and $\Psi^{\star}=\Psi $.
In chiral  basis

 \beq
\gamma^0=
\left(\begin{array}{cc}
0 & 1 \\
 1& 0\\
\end{array}
\right ),\,
\gamma^i=
\left(\begin{array}{cc}
0&  \sigma^i \\
-\sigma^i&0\\
\end{array}
\right ),\,
\gamma^5=
\left(\begin{array}{cc}
1 & 0 \\
0 & -1\\
\end{array}
\right ),\,
C=i{\gamma}^2=
\left(\begin{array}{cc}
0& {\epsilon}^{\alpha\beta} \\
-{\epsilon}^{\alpha\beta} & 0\\
\end{array}
\right ).
\eeq
Note that the Majorana spinor can be expressed in terms of 
Weyl spinors as

\beq 
\Psi_M=
\left(
\begin{array}{c}
{\vartheta}_a \\
-{\epsilon}^{\alpha\beta}{\bar \vartheta}^{\dot a}\\
\end{array}
\right )
\eeq

 Theory of a massless neutrino in $D=3+1$ can be described either a theory of a massless Weyl fermion or as a theory of a massless four-component Majorana fermion.  
  In Majorana case the neutrinos  have the following
 mass term:
 \be 
 m_M(\nu) {\bar{\Psi}}_M {\Psi}_M=m_M(\nu)({\vartheta}^a{\vartheta}_a +
 {\bar \vartheta}_{\dot a}{\bar \vartheta}^{\dot a}).
 \ee 
 It is very important to establish  a link between masses of three families.
To get some additional  relations between the masses  of three Majorana neutrino species one can try to study  
the possible $D=6$ space-time global symmetries how it was done in $D=3+1$ Minkowski space-time. 
This symmetry could give us a new relation which could be  a signal from the extra dimension space-time. Our goal is to check the logic chain of  the discovery the space-time   nature, of spin, anti-particles and  
the three families (?). 
After detail study the space-time properties of neutrinos in $D=3+1$ it is 
tempting  to consider   all three neutrinos 
in one $D=6$ "ternary six-dimensional field"  

\beq 
(\psi_{{\nu}_e}, \psi_{{\nu}_{\mu}},\psi_{{\nu}_{\tau}})
\rightarrow \Psi_ {3\nu}.
\eeq
 Taking into account already known six space-time degrees of freedom of the three
 Lorentz-Majorana neutrinos  one can consider them in the $D=6$ space-time.
 In this field   each "internal" 2-component   spinor  bears   the information about its  $D=(3,1)$  space-time  structure.  In this $D=6$-space apart from ordinary complex conjugation it is possible to embed the procedure of  ternary  conjugation using the  $C_3$-ternary complex numbers with   the  generator  $q$  satisfying to the following
conditions:
\be 
q^3=1.
\ee

We remind that if for the cyclic group $\mathit{C}_{2}$ there are two
conjugation classes, $1$ and $i$ and two one-dimensional
irreducible representations:

\beq
\begin{array}{c|cc|c}
\mathit{C}_{2} & 1 & i &  \\ \hline
R^{(1)} & 1 & 1 & z \\
R^{(2)} & 1 & -1 & \bar{z}
\end{array}
\eeq
 the cyclic group $\mathit{C}_{3}$ already has three conjugation classes,
$q_{0}$, $q$ and $q^{2}$, and, respectively, three one dimensional
irreducible representations, ${R}^{(i)}$, $i=1,2,3$. We write down
the table of their characters, $\xi _{l}^{(i)}$:

\beq
\left (
\begin{array}{c|ccc}
    -         & 1 & q    & q^2  \\\hline
    \xi^{(1)} & 1 & 1    &  1   \\
    \xi^{2)}  & 1 & j    & j^2  \\
    \xi^{(3)} & 1 & j^2  & j    \\
\end{array}
\right)
\eeq

for ${ C}_3$  ( $j_3=\equiv j=\exp \{2 \pi /3 \}$).

This table defines the operation of ternary complex conjugations:
\beq   
\tilde {q}&=&j q, \quad \tilde {{\tilde {q}}}=j^2q, \quad
\tilde {\tilde {\tilde {q}}}=q, \nonumber\\ 
{\tilde {q}}^2&=&j^2 {q}^2, \quad {\tilde{\tilde {q}}}^2 =j q^2 \\
  j&=&\exp{(\frac{2\pi\,i}{3})}, \qquad 1+j+j^2=0.\\
\eeq 

So, the main idea is to use the cyclic groups ${ C}^n$
(and new algebras/symmetries)
to find the new geometrical "irreducible"' substructures in
${ R}^n$ spaces, which are not the  consequences of the simple direct n-multiplication
of the known structures of Euclidean ${R}^2$ space which could help to build the $D=6$ extension of the Lorentz group.

Than one can introduce the "family" ternary quantum 
charge numbers  which should be conserved by a new ternary Abelian group.
If this symmetry in $D=6$ exists it means that in this space-time
 can  exist a new "light" expanding with a velocity faster than electromagnetic light?! The structure of this ternary Abelian symmetry we discuss in the next section.
 
Then  3-neutrino field can be represented as 
\beq 
\Psi=
\left (
\begin{array}{ c}
\vartheta\\
\tilde {\eta}\\
\tilde {\tilde {\xi}}\\
\end{array}
\right)
\eeq
 
The action of the new charge operator on the field 
can be constructed with the operator expressed in    $6\times 6$- matrix  form:

\begin{eqnarray}  
\Gamma^1=
\left(
\begin{array}{ccc}
 {0}& \sigma^0 & {0}\\
{0}& {0}&  \sigma^0  \\
\sigma^0 & {0}& {0}\\
\end{array} 
\right),
{(\Gamma^1)}^2=
\left(
\begin{array}{ccc}
{0}& {0}&   \sigma^0  \\
\sigma^0 &  {0} &  {0} \\
 {0} & \sigma^0 &  {0}\\
\end{array} 
\right),
{(\Gamma^1)}^3=
\left(
\begin{array}{ccc}
 \sigma^0 &{0}      &  {0}\\
{0}       &\sigma^0 &  {0}  \\
 {0}      & {0}     &  \sigma^0 \\
\end{array} 
\right).
\end{eqnarray} 
The consequent actions of this operator are
\beq   
&&C^{g}\Psi = \Gamma^1 \tilde {\Psi} = \Psi^{g},\nonumber\\
&&C^{g}C^{g}\Psi=C^{gg}\Psi=\Psi^{gg} \nonumber\\   
&&C^{g}C^{g}C^{g}\Psi=C^{ggg}\Psi=\Psi^{ggg}=\Psi,\nonumber\\
\eeq
 where
 \beq 
 \Psi^{g}=
 \left (
 \begin{array}{c}
 \tilde{\vartheta}\\
\tilde {\tilde {\eta}}\\
 {\xi}\\
\end{array}
\right)
\eeq
 and 
 \beq 
  \Psi^{gg}=
 \left (
 \begin{array}{c}
 \tilde {\tilde{\vartheta}}\\
 {\eta}\\
 \tilde {\xi}\\
\end{array}
\right)
\eeq

So we presented the 3-neutrino  specie with respect to a global symmetry
in $D=6$ as the ternary Dirac complex fermion each component we considered as Majorana with respect to the  Lorentz symmetry. 
The action  in $D(p+q=6)$ space-time  of the  the discrete possible $[C{g}P^{p}T^{q}]_6$-invariance   for three neutrino species 
 could  lead to the  mass relation as   it  was in $D=(1+3)$ space-time with application  $CPT$ theorem:

\be
m_M(\nu_{e} )= m_M( \nu_{\mu})=m_M(\nu_{\tau})\qquad Ternary \, mass \, relation  \\
\ee   
The mechanism of oscillation could be done due to a hypothetical
interaction existing in $D=6$  closely reminding the mechanism of transitions of the $K^0-{\bar{K}}^0$ mesons.
As $CPT$ theorem is acting only in $D=4$ these oscillations can go with $CPT$-violation. All these consequences could be checked on experiments.
Thus this symmetry could send  a signal from the extra dimension space-time.

Moreover in this case one can consider the chiral ternary symmetry, which is acting in all $D=6$ space-time for $U(1_{em})$ sterille particles. In this case the ternary chiral symmetry  forbids all bilinear Majorana 
 mass forms. Thus we could come to the consequence that all Majorana-Lorentz  masses  are equal zero?!  It seems the same result could be also valid for Dirac-Lorenz mass bilinear terms.
In general the negligibility  of neutrino masses with respect to the 
\EWSB ($m_{\nu}/M_{EW}\sim 10^{-9}$) can confirm that the extra-dimensional dynamics could be very important to understand this very intriguing question.  For example, one can represent the existence
the hot dark matter what emits  a new "light" expending much faster  than light?  The massless of the particles could lead to a new symmetry....
There is the question with charged fermions, quarks and charged leptons.
The mechanism of production of the masses is connected to the \EWSB and 
as one can  see the mass spectrum  is confined by  this scale.
The charged particles are living in the $D=3+1$ space-time and 
the ternary symmetry is breaking in this part of $D=6$.

\section{  Towards a  ternary theory of neutrino.}

 Among unresolved problems of Standard Model in a special way there are questions on an origin and a role of 3 types neutrinos in the Universe and, at last, their special characteristics. On the one hand, neutrino participates only in weak interaction and consequently possesses deeply getting property at substance passage. On the other hand, neutrinos, unlike others fermion components of the Standard Model, have very small mass. This fact in itself is the extremely mysterious problem, but it leads to that neutrinos are long-living.

The third  possible important property which can have neutrinos, is connected with their possible space-time   nature in $D=4$ or in $D=6$. 
 Neutrinos while are unique representatives of a matter known to us from the Standard Model which could be connected with the physical phenomena, originating from more extensive multidimensional geometry of our Universe filled with more fundamental dark matter?
According to the Standard Cosmological Model and last astrophysical data, the dark matter makes about $22 \%$ from all structure of substance of the Universe. The contribution of dark energy makes $74 \%$, and the matter observed in the form of planets, stars, galaxies, intergalactic gas makes all about $4 \%$. Undoubtedly that these estimations are dependent from  modelling are  represent our understanding of world around at present.
 As a result we come to a conclusion that already observable quark-lepton the matter spectrum, a number of other researches of already spent experiments within the limits of Standard Model together with the resulted data received from astrophysical researches, specify in existence of absolutely new properties of structure of the vacuum connected with new space-time geometry of the Universe.

Electroweak symmetry in  \SM  is broken to $ U (1_{em})$-symmetry using  Higgs mechanism.  
But there was represented the very interesting direction, in which
violation of electroweak symmetry is connected with additional global dimensions. So the problem of the neutrino masses could be decided at the assumption that ours 
$D=3+1$ world is a part of 6-dimensional space-time in which some symmetries forbid the "usual" mass terms 
( "threshold effect from $D=6\,  \rightarrow \,D=4$")?!

Also, it is necessary to consider still observable properties of exotic global symmetry 3 quark – lepton generations which because of considerable distinction of the masses should be strongly broken. Quarks and leptons enter into Standard Model in the form of 3 generations which except for their mass spectrum can be considered as identical repetitions of one generation. Distinction of the masses of quarks of 3 generations is shown in occurrence of their mixing which has been measured on experiment in electroweak quark  decays. Despite intensive and long experimental searches, a question on existence gauge interactions between the fermions of the various generations till now remains opened and represents the next riddle of the nature. There have done a lot of efforts to solve this problems by searching  for local gauge family symmetries, for example in \SS\, there were suggested two variants with 3-families with 
gauge symmetry $U(1)\times U(1)\times U(1)$ in \cite{AEHN}
and $3+1$ families with $SU(3)\times U(1)$ gauge symmetry  in \cite{VOL}.
Many experimental searching the rare processes with lepton number violation did not give any success. So there  appeared the question that the family problem could be connected to the new space-time symmetries having an origin to the extra dimensions.  
So our  way to try to solve the family problem by geometrical way through the new space-time symmetries and at first we  concentrated on three neutrino problem.  The charged quarks and leptons have considerably different properties.
To solve  this problem we decided to use   the ternary symmetries which are connected to the ternary complex numbers and, correspondingly,  to the complexification of $R^{3n}$ -Euclidean spaces.
The matrices Pauli $\sigma_{\mu}$ have been used in non-relativistic (Galilei) quantum mechanics of theory of electron  for description of spin:
\beq 
&& \sigma_i \sigma_j+\sigma_j\sigma_i=\delta_{ij}  \nonumber\\
&& \sigma_i^2=\sigma_0, \qquad 
\sigma_i\sigma_j\sigma_k=i\epsilon_{ijk}\sigma_k  \nonumber\\
\eeq
These matrices can be used to construct the $4\times 4$   $\gamma^{\mu}$, Dirac matrices which satisfy to the following relation:
\beq
\gamma_{\mu}\gamma_{\nu}+\gamma_{\mu}\gamma_{\nu}=2 g_{\mu\nu},
\eeq
where $g^{\mu\nu}=diag(1,-1,-1,-1)$.
The matrices Pauli have been used in non-relativistic (Galilei) quantum mechanics
of theory of electron what used the 2-dimensional spinors for 
description of spin. The Dirac matrices $2^2\times 2^2$ have been used in relativistic Dirac equation of electron what expends the notion of spin in into $D4$-space-time and gave  the origin of a new anti-fermion matter.
If one uses this binary Clifford  algebra for $D=6$ space-time he should introduce the  $2^3\times 2^3$ Dirac matrices what can build from
Pauli matrices by the next rules:
\beq  
\begin{array}{ccc}
&\gamma_1^{(3)}=\sigma_1\otimes \sigma_0\otimes\sigma_0
&\gamma_2^{(3)}=\sigma_2\otimes \sigma_0\otimes\sigma_0 \nonumber\\
&\gamma_3^{(3)}=\sigma_3\otimes \sigma_1\otimes\sigma_0
&\gamma_4^{(3)}=\sigma_3\otimes \sigma_2\otimes\sigma_0 \nonumber\\
&\gamma_5^{(3)}=\sigma_3\otimes \sigma_3\otimes\sigma_1
&\gamma_6^{(3)}=\sigma_3\otimes \sigma_3\otimes\sigma_2\nonumber\\
\end{array}
\eeq
For the consideration the $D=2^{(2k+1)}$ dimensions one should introduces the analogue of $\sigma_3$ , $\gamma_5$ - matrices.

If the origin of Lorentz symmetry, Dirac equations, $\gamma^{(k)}$-matrices and etc started from geometry of the plane- Pythagoras theorem
and symmetries of the circle, hyperbola, ordinary complex analysis.
 In addition one can  study new symmetries, new manifolds, new Pythagoras theorem and etc  in $R^3$. In  $R^3$ there  could be some new  structures what you cannot get from the simple  composition of the $R^2$-plane  and $R^1$-line.
 It reminds the problem of three body interactions where to get correct result one should take into account the irreducible triple interaction. 
 The  extension of  the $D=3+1$-space-time  till $D=6$ by Dirac procedure 
gives the doubling increase of the fermion spectrum . To get the tripling  
 increase of the fermions (three families) one should another way,
 {it i.e.} "to study the  theory of the number 3": the complexification
 of $R^3$, the cubic Pythagoras theorem, the new structures of $R^3$
 and new symmetries and etc \cite{Appel},\cite{Dev},{\cite{Fleury}, \cite{Kern},\cite{Rausch},\cite{LRV},\cite{DV}, \cite{Volk}: 
 
 Let start from the ternary complex numbers:
\beq
z&=&x_0 q_0 + x_1 q + x_2 q^2,  \nonumber\\
\tilde{z} &=& x_0 q_0 + j x_1 q + j^2x_2 q^2, \nonumber\\
\tilde  {\tilde {z}}&=& x_0 q_0 + x_1 j^2 q + x_2 j  q^2. \nonumber\\
\eeq
Now one can build the following cubic form:
\begin{eqnarray}
<z>^3&=&z {\tilde z} {\tilde {\tilde z}}\nonumber\\
<z_1z_2>^3&=&<z_1>^3<z_2>^3  \nonumber\\
\end{eqnarray}

  The following exponent expression 
\be  U_T(1)=\exp{(q \phi_1 + q^2\phi_2) } \nonumber\\
\ee
with two conjugation operations 
\beq 
&&\tilde{U}_T(1)=\exp{(jq \phi_1+j^2 q^2 \phi_2)} \nonumber\\
&&\tilde {\tilde {U}}_T(1)=\exp{(j^2\phi+jq^2 \phi_2)}.\nonumber\\
\eeq
produces the Abelian two-parameter ($\phi_i$) group.
This exponent  has very important property:
\be  
U_T \cdot {\tilde {U}}_T \cdot {\tilde {\tilde {U}}}_T=1 
\ee 
what can be expanded for n-dimensional   ternary  unitary operators:
\be 
U_T \cdot  {U_T}^{\dagger} \cdot {U_T}^{\dagger\dagger}=1.
\ee
The Abelian unitary $U_T$-group can be expanded in any $n$
what is defined by $2n^2-1$-parameters (n>1). 
 
 The unit form determines the $U_T(1)$ invariant two-dimensional surface:
 \be 
z \tilde {z} \tilde {\tilde {z}}=
x_0^3+x_1^3+x_2^3-3x_0x_1x_3=1
\ee
This cubic surface(spherical  parabaloid)  is a ternary analogue of the $S^1$- circle defined by units complex numbers : $z \cdot \bar {z}=1$.

Similarly one can build the triple  $U_T(1)$-unitary invariant form in the space of the fields:
\be 
\psi\tilde {\psi} \tilde{\tilde {\psi}} 
\ee
Thus the $U_T(1)$- transformations could be linked to the space symmetry of the cubic surface (spherical parabaloid) and to the invariant production of the norm functions.

The unit ternary complex numbers $||q||=1$ can be realized
in the set of the $3 \times 3$ matrices-nonions
 \cite{Fleury},\cite{Kern}, \cite{Rausch},\cite{ LRV},\cite{Volk}- the three dimensional extension of Pauli matrices  having the some common properties:
\beq
 Q_a^3&=&1Q,\qquad a=0,1,\cdots, 8,\nonumber\\
Q_{3+i}&=&Q_{i}^2={Q_i}^{\dagger},\quad i=1,2,3,\nonumber\\
Q_{i}&=&Q_{3+i}^2={Q_{3+i}}^{\dagger} \quad , i=1,2,3\nonumber\\
&& Q_7^2=Q_8,\qquad Q_8^2=Q_7 \nonumber\\
\eeq
See also  the full table of multiplication  producing the 
 ternary analogue of quaternions in \cite{DV}.

\beq
Q_1 = 
\left( 
\begin{array}{ccc}
0 & 1  & 0 \\
0 & 0  &  1 \\
1  & 0  &  0 \\
\end{array}
\right ),  \qquad 
Q_2 = 
\left (
\begin{array}{ccc}
0 & 1  & 0 \\
0 & 0  &  j \\
j^2  & 0  &  0 \\
\end{array}
\right ), \qquad 
Q_3 = 
\left (
\begin{array}{ccc}
0 & 1  & 0 \\
0 & 0  &  j^2 \\
j  & 0  &  0 \\
\end{array}
\right )   \nonumber\\
\eeq

\beq
Q_4 = \left( 
\begin{array}{ccc}
0 & 0  & 1 \\
1 & 0  & 0 \\
0 & 1  &  0 \\
\end{array}
\right ),  
\qquad Q_5 = \left (
\begin{array}{ccc}
0 & 0  & j \\
1 & 0  & 0 \\
0  & j^2  &  0 \\
\end{array}
\right ),
 \qquad Q_6 = \left (
\begin{array}{ccc}
0 & 0  & j^2 \\
1 & 0  &  0 \\
0 & j & 0 \\
\end{array}
\right )   \\
\eeq

\beq
Q_7  =\left( 
\begin{array}{ccc}
1 & 0  & 0 \\
0 & j  & 0 \\
0 & 0  & j^2 \\
\end{array}
\right ),  
\qquad Q_8 = \left (
\begin{array}{ccc}
1 & 0  & 0  \\
0 & j^2  & 0 \\
0  & 0  &  j \\
\end{array}
\right ),
 \qquad Q_0 = \left (
\begin{array}{ccc}
1 & 0  & 0 \\
0 & 1  &  0 \\
0 & 0 & 1 \\
\end{array}
\right )   \\
\eeq

The matrices of the  I-st  and II-nd sets satisfy to some  remarkable relations:

\beq
&&Q_iQ_jQ_k + Q_jQ_kQ_i+Q_kQ_iQ_j
= 3 j^2\eta_{ijk} Q_0\nonumber\\
&&Q_{3+i}Q_{3+j}Q_{3+k} + Q_{3+j}Q_{3+k}Q_{3+i}+Q_{3+k}Q_{3+i}Q_{3+j}
= 3j \eta_{ijk} Q_0\nonumber\\
\eeq

with
\begin{eqnarray}
&&\eta_{111}=\eta_{222}=\eta_{333}=1 \nonumber\\
&&\eta_{123}=\eta_{231}=\eta_{312}=j \nonumber\\
&&\eta_{321}=\eta_{213}=\eta_{132}=j^2 \nonumber\\
\end{eqnarray}
where $j=\exp (2 \pi/3)$.

The surface

\begin{equation}
x_0^3+x_1^3+x_2^3-3\,x_0x_1x_2=1
\end{equation} 
after changing coordinate system can get the famous expression.
 Introduce $u_0=x_0+x_1+x_2$ and parametrise a
point on the circle of radius $r^2=u_1^2+u_2^2$ around the trisectrice by its polar coordinates $(r,u_1=r \cos\theta, u_2=r \sin \theta)$. The surface
in these coordinates
can be reduced to the well known  spheric parabaloid:
\begin{equation}
u_0 (u_1^2+u_2^2)=1
\end{equation}

From this cubic surface one can also define the tetrahedron Pythagoras  theorem:

\begin{eqnarray}
 S_A^3+S_B^3+S_B^3- 3 S_AS_BS_C= S_D^3,
\end{eqnarray}
what gives in the tetrahedron  the link between the areas
of the  $S_{A,B,C,D}$ four triangle faces.
This theorem can be easily extend to all $D=4,5,6,...$-dimensional cases!
 
 From the   exponent expression one can find the space structures of this ternary unitary group:
\beq  
U_T=\exp{(q \phi_1 + q^2\phi_2) }=\exp{(q-q^2)\alpha)} \cdot \exp{(q+q^2)\beta))}\rightarrow SOP(2,1) \nonumber\\
\eeq
The $SO(2^{\prime}){\hat{\times}}{t} P(2,1)$ group of transformations  
produced the one-parameter three-dimensional  orthogonal group and "projective" parabolic group.
Remind that the $SO(2^{\prime})$-symmetry is acting in $R^3$-spaces,
where the  general orthogonal group $SO(3)$ has three parameters.
The "projective" group  $P(1,2)$ defines the set of parabolas in $R^3$-spaces. 

Considering the "real" case
\beq 
U_T \rightarrow SOP(2,1)
\eeq 
 The subgroup  $SO(2^{\prime})$ can be expressed through the $3\times 3$ matrix \cite{Fleury},\cite{Kern}, \cite{Rausch},\cite{DV}, \cite{ LRV},\cite{Volk} .

The subgroup  $SO(2^{\prime})$ can be expressed through the $3\times 3$ matrix:

\begin{eqnarray}
SO(2^{\prime})= \exp \{\alpha (q_1-q_1^2)\}=
\left (
\begin{array}{ccc}
 c_0& s_0 &t_0  \\
 t_0& c_0 &s_0  \\
 s_0& t_0 &c_0  \\
\end{array}
\right ),
\end{eqnarray}            \nonumber\\
where we have the particular choice for the functions, $c$, $s$,
$t$ giving the determinant equal to one:
\begin{eqnarray}
&&c_0=\frac{1}{3}(1+2 cos (\phi)) \nonumber\\
&&s_0=\frac{1}{3}(1+2 cos (\phi +\frac{2\pi}{3})), \nonumber\\
&&t_0 =(\frac{1}{3}1+2 cos (\phi - \frac{2\pi}{3})). \nonumber\\
\end{eqnarray}

where $j-j^2=\sqrt{3}{\bf i}$, $\phi=\sqrt{3}\alpha$.

 Apart from that the    $SO(2^{\prime})$ transformations preserve
 the cubic form they also
satisfy to binary condition $O gO^t=O^tg O=1$, preserving the quadratic metric $g_{ij}=diag(1,1,1)$ what is equivalent to the
additional  two restrictions for the ordinary $3\times 3$ 
orthogonal transformations:
\begin{eqnarray}
&& c_0^2+s_0^2+t_0^2=1 \nonumber\\
&& c_0s_0+s_0t_0+t_0c_0=0,  \nonumber\\
\end{eqnarray}
what in our case can be easily checked. Thus in the case  
$\alpha=[(\phi_1+\phi_2)/2] =-\beta=-[(\phi_1-\phi_2)/2]$ =
the ternary symmetry gives the orthogonal symmetry in $R^3$  with additional ternary restriction.

In the case  $\alpha=\beta $, the ternary symmetry coincides
with the other binary symmetry preserving the quadratic metrics:
\beq
g_{ij}=diag(2,-1,-1), \qquad P^t\cdot \hat g \cdot P=\hat g. 
\eeq
\begin{eqnarray}
P= \exp \{\alpha (q_1+q_1^2)\}
= \left (
\begin{array}{ccc}
    c_{+}  &   s_{+}   &  s_{+}   \\
    s_{+}  &   c_{+}    &  s_{+}     \\
    s_{+}  &   s_{+}    &  c_{+}        \\
\end{array}
\right ),
\end{eqnarray}
where $ c_{+}= \frac{1}{3}(e^{\{2\alpha\}}+  2 e^{\{-\alpha\}} )$,
      $ s_{+}= \frac{1}{3}(e^{\{2\alpha\}}-    e^{\{-\alpha\}} )$
and the cubic equation reduces to the next form:
\begin{eqnarray}
 c_{+}^3+ s_{+}^3+ t_{+}^3-3 c_{+} s_{+} t_{+}=( c_{+}- s_{+})^2( c_{+}+2 s_{+})=1.
\end{eqnarray}

Note that apart from the remarkable group properties  the matrices 
\beq= O_T=
\left (
\begin{array}{ccc}
c & s & t \\
t & c & s\\
s & t & c \\
\end{array}
\right ) , 
\qquad det (O_t)=c^3+s^3+t^3-3cst,
\eeq
satisfy to the next  equality:

\begin{eqnarray}
O_T \cdot {O_T}^J \cdot {O_T}^{JJ}=
c^3+s^3+t^3-3cst \cdot \hat 1,
\end{eqnarray}
where  we introduce the following operations of conjugations:
\begin{eqnarray}
{O_T}^J=
\left (
\begin{array}{ccc}
c & js & j^2t \\
j^2t & c & js\\
js & j^2t & c \\
\end{array}
\right ) \qquad 
{O_T}^{JJ}=
\left (
\begin{array}{ccc}
c & j^2s & jt \\
jt & c & j^2s\\
j^2s & jt & c \\
\end{array}
\right ) \nonumber\\
\end{eqnarray}

Thus, the two parametric ternary  group $SOP(1,2)$ preserving the cubic form reduces exactly to
two known binary symmetries, $\alpha=-\beta$ and  $\alpha=\beta$,
but  in which these two binary symmetry are unified by non-trivial way.

The extension of group $SOP(2,1)$ from $R^3$ space to the    $R^n$ spaces ($n>3$)  should have the $2\times C^n_3$-parameters and construction 
the transformation in matrix forms  is following to the known extensions of usual orthogonal groups which have the $C^n_2$-parameters.

\section{ Preliminary conclusions  from  $D=(3+1=4)\subset   D=(p+q=6)$. }

The three species of  neutrinos could have a ternary complex structure which could  directly related to the origin of three quark-lepton families.  In this approach the family symmetry can have the space-time origin as we already have met with similar  question of the origin matter-antimatter in Dirac relativistic theory of electron. 
Since there are three  quark-lepton  families  the procedure of Dirac cannot use any more. In his approach we could hope only to get
in every stage the doubling of fermion states: $2^{\frac{d}{2}}$. To get tripling or to get three families we should include in consideration new properties of the space-time. 
Before all our progress was connected to theory the well known properties of the two-dimensional  Euclidean, Lobachevski and Riemann surfaces   and its extensions to the n-dimensional cases. In this case all 
symmetries and  algebra are based on the binary structures, for example, the possible invariant quadratic metrics $g^{\mu\nu}$ has the signature $\{(+....+)_p,(-...-)_q\}$.  So if in the old approach there were used the Pythagoras theorem for triangle, the complex analysis for the plane
but in our case we need also to include  the "irreducible" 
structures going only from  $R^3$ spaces, where exists the  tetrahedron
extension of Pythagoras theorem, the ternary complex numbers and the ternary complex analysis with the cubic  differential Laplace equations \cite{Fleury},\cite{Rausch},\cite{Kern},\cite{DV}, \cite{LRV},\cite{Volk}.
There is very big difference between the charge matter fermions and 
neutrinos what can be see from the structure of Lorentz group representations: Dirac, Weyl chiral and Majorana fermions.
Suggesting the two dimensional structure of neutrino field ( Majorana or Majorana-Weyl we came to the consideration the  ($D=2(spin)\times 3(fam)=6$-space-time. It gives us a chance to include in our ambient space-time just two additional extra dimensions.
To realize this project we studied the $3\times 3$ matrices- nonions-
the three dimensional extension of the 4-Pauli matrices.
 The algebraic structure of nonions is much complicate than the algebraic structure of Pauli matrices.  The ternary complex analysis allowed us to
 discover the Abelian groups  preserving the cubic surfaces and the n-dimensional extensions of them to non Abelian case.
 To observe the new extra dimensions it was suggested to make the neutrino experiments to measure the neutrino velocity which as was predicted in \cite{AV} is more than light. 
Such scenario with some fundamental boost-velocities could give a push to go beyond
the standard Big Bang model in the time before the SU(2)×U(1) phase and could explain
the horizon, ﬂatness problems. This scenario is diﬀerent with respect to the other scenario
of the varying speed of light \cite{Mof} in spite of the similar problems to construct such
mechanisms.
The restriction of sterility for neutrino in our conjecture means that with the electromagnetic charged particles to observe new boost (new topological circuit) at the now
available energies is very difficult or may be impossible now.
 For U(1)em charged particles “getting” to the new SU(2) × U(1) vacuum
structure could be a threshold effect like as Vavilov-Cherenkov eﬀect with emitting of lot of energy \cite{AV}. May be, this threshold could be linked to the scale of electroweak symmetry violation.

 The other signatures which could confirm the result \cite{OPERA} are the searching for the processes going with violation  $CPT$-invariance and $Q_{em}$ conservation. For example, it is important to search for the $CPT$ invariance violation in neutrino
 oscillations \cite{MINOS2007}.

The new ternary structure structure of Majorana neutrino
can lead to discovery the new interactions linked the  electromagnetic charge matter to the dark matter.
  There is very important the  progress in making the  new 
experiments  to check is  neutrino of Majorana-Weyl  nature or not.
The other possibility to check the Majorana-Weyl  nature of neutrino can
be related with the proton   decay problem where
the proton non-stability problem could be solved also in the
$electron$, $\mu-$, $\tau-$ lepton and $b-$ meson  rare decay
experiments \cite{Paris} which are going   with $Q_{em}$-conservation
law  violation.

We started our discussion from the exceptional properties of the Lorenz group structure
-Dirac-Weyl-Majorana neutrinos- which may be have non-Majorana structure 
with respect to a new D=6 space-time symmetry- and step by step
want to come to an idea of  an existence of completely new physics
at energy scale $O(1-10) TeV$ which is related to the D=6 space-time geometry \cite{Paris} and may be with origin of the charge fermions in 
\SM .   

The new ternary structure could be linked to the
some global dark symmetries in \SM, for example:

\beq
N(colors)\,=\,N(families)\,=N(space\, dim.)\,=\, 3.
 \nonumber\\
\eeq

Expected new experimental and theoretical consequences from carrying out further neutrino experiments cause so huge interest in scientific and public circles of which wasn't since opening of the Special theory of a relativity in the beginning of 20th century. Multidimensional generalization of the special theory of a relativity will result and in respective generalization of the General Theory of the Relativity and will bring very important attention to the question on speed of distribution of gravitational waves

The expected results  of the neutrino measurements of velocity could be also  depend on the possible signature $(p+q)=6, p\geq 3, q\geq 1$  of two extra dimensions  in D=6- space-time:
\beq
\begin{array}{ccc}\hline
 D=p+q=6  &  space dim. &  time dim.\\\hline
R^6_{5,1} &   p=5  & q=1\\\hline
 R^6_{4,2}  &  p=4  &q=2\\\hline
R^6_{3,3}  & p=3   & q=3 \\\hline
\end{array}
\eeq
At \cite{AV} we suggested to check the possible effect $V_{neut} >c$
in three  types neutrino experiments:
\begin{itemize}
\item{1.The experiments with a long base: MINOS, OPERA, K2K}\\ 
\item{2.The experiments with the short base}\\
\item{3.Beam-dump experiments}\\
\end{itemize}
The main question for us was not only to fix the effect but to understand the dynamics of the possible  effect,  for example,   dependence on the neutrino energy or on the mass of the neutrino sources,  dependence on the neutrino species and Majorana-Weyl structure,
how this possible effect could depend  energy-distance-direction-time?  

 At the end of this article we would like to stress that 
 to get correct results there should be done a lot of  experimental and theoretical investigations in future.

\vspace{0.5cm}
\begin{enumerate}
\item{The Majorana-Weyl-Dirac structure of neutrinos 
in $D+(3+1) \rightarrow D=(p+q)$  space-time}\\
\item{Complexification of $R^{3},R^{4},R^{5}, R^{6}$ and the corresponding Abelian symmetries}\\
\item{ Classification of   ternary non- Abelian symmetries}
\item{The new space-time structure of neutrinos in cosmology with extra dimensions $D>4$}\\
\item{A link neutrino with dark matter}\\
\item{Dark ternar external and internal symmetries  symmetries in the SM}\\
\item{Poincar$\acute{e}$ duality: $CPT$-invariance and
$Q_{em}$-conservation}\\
\item{The link between masses of quarks/charged leptons and electro-weak symmetry breaking}
\item{New interactions and proton/electron non-stability}
\end{enumerate}

\section{Acknowledgements}
I thank Alexey Erikalov and Lev Lipatov for support and valuable discussions during many years on working on this subject.
Sometimes it was very dangerous  to work on so fundamental
questions related to \STR\,. Our life with Vladimir Ammosov confirmed
this. 
So many negative emotions we have met during our long way of working on this problem $v_{\nu}>c$.    Therefore I would like to express my acknowledgements to Antonino  Zichichi, director of World Laboratory
in CERN,  where I have got the possibilities to work and to discuss with many my colleagues.  I thank    John Ellis and  Dimitri Nanopoulos for very important idea of the dynamical gravitation mechanism of  breaking the space-time symmetries $ SO(3,1) \rightarrow SO(2)\times SO(1,1)$\cite {Ellis}
and its possible extension to the  $D=6$ case: $SO(4,2)\rightarrow SO(3,1)\times SO(1,1)$ ( new boost). There was very interesting discussion with Robert  Yamaleev about the ideas that the three families could be considered as $f \rightarrow  f^{\star} \rightarrow f^{\star\star}$.
Also I thank  Franco Anselmo, Ugo Aglietti, Guennady Danilov, Alexandr Dubrovskyi, Oleg Fedin, Semen Gerstein,  Alexandr Kisselev, Michail Konoplya, Andrey Kuznetcov, Dimitri Leonenko,  
Vladimir Obraztcov, Zoja Neradovskaja, Michel Rausch de Traubenberg,  Serguei Sadovski, 
Vladimir  Samoylenko, 
Anatoly Sokolov, Michail Tchernetcov,  Sergei Volkoff for many valuable discussions and advices.
I would like to thank  George  Bitsadze for support, for discussions  and reading  manuscript.

\end{document}